\documentstyle[12pt]{article}

\newcommand{\beq}{\begin{equation}}
\newcommand{\eeq}{\end{equation}}
\newcommand{\bea}{\begin{eqnarray}}
\newcommand{\eea}{\end{eqnarray}}

\renewcommand{\a}{\alpha}

\newcommand{\kt}{k_\perp}

\newcommand{\ra}{\rightarrow}

\newcommand{\s}{\hat s}
\newcommand{\th}{\hat t}
\renewcommand{\o}{\cal O}
\newcommand{\nn}{\nonumber}
\newcommand{\un}{\underline}

\begin{document}
\topskip 2cm 
\begin{titlepage}

\hspace*{\fill}\parbox[t]{4cm}{EDINBURGH 97/10\\ July 1997}

\vspace{2cm}

\begin{center}
{\large\bf BFKL in forward jet production}\\
\vspace{1.5cm}
{\large Vittorio Del Duca} \\
\vspace{.5cm}
{\sl Particle Physics Theory Group,\,
Dept. of Physics and Astronomy\\ University of Edinburgh,\,
Edinburgh EH9 3JZ, Scotland, UK}\\
\vspace{1.5cm}
\vfil
\begin{abstract}
In this talk I consider dijet production at large rapidity 
intervals in hadron collisions and forward-jet production in DIS
as candidate signatures of a BFKL evolution. The state of the art
on the measurements, and on the BFKL-motivated phenomenological
analyses with emphasis on the different approximations involved, 
is reviewed.
\end{abstract}

\vspace{2cm}

{\sl To appear in the Proceedings of the\\ XIIth Hadron Collider Physics
Symposium\\ Stony Brook, NY, June 1997}

\end{center}

\end{titlepage}

\section{Dijet production at large rapidity intervals}
\label{sec:uno}
\subsection{The BFKL ladder}
\label{sec:unozero}

About ten years have past since Mueller and Navelet \cite{mn} 
have proposed to test in dijet production at hadron colliders
the BFKL theory \cite{bal}, modelling
strong-interaction processes with two large and disparate scales,
and my goal here is to describe how their original proposal has
evolved.
First, I shall briefly summarize what the BFKL theory is about:
in the limit of center-of-mass energy much greater than
the momentum transfer, $\s\gg|\th|$, any scattering process 
is dominated by gluon exchange in the cross channel, which 
occurs at ${\o}(\a_s^2)$; thus, if we take parton-parton scattering 
as a paradigm process, the functional form of the amplitudes
for gluon-gluon, gluon-quark or quark-quark scattering is the
same; they differ only for the color strength in the parton-production
vertices\footnote{This may be used as a diagnostic tool for 
discriminating between different dynamical models for parton-parton 
scattering. Namely, in the measurement of dijet angular distributions,
models which feature gluon exchange in the cross channel, like QCD, 
predict a parton cross section, and therefore a dijet angular 
distribution, which doesn't fall off as $\s/|\th|$ grows \cite{EKS}, 
\cite{dijet}, while models featuring
contact-term interactions don't have gluon exchange in the cross channel,
and thus predict the dijet angular distribution to fall off as 
$\s/|\th|$ grows \cite{cont}.}. To higher orders, we may resum the 
contribution of the radiative corrections to parton-parton
scattering to leading logarithmic (LL) accuracy, in
$\ln(\s/|\th|)$, through the BFKL equation, i.e. a two-dimensional
integral equation which describes the evolution in transverse momentum
space and moment space of the gluon propagator exchanged in the cross
channel,
\beq
\omega\, f_{\omega}(k_a,k_b)\, =
{1\over 2}\,\delta^2(k_a-k_b)\, +\, {\a_s N_c\over \pi^2} 
\int {d^2k_{\perp}\over k_{\perp}^2}\,
K(k_a,k_b,k)\, ,\label{bfklb}
\eeq
with $N_c=3$ the number of colors, $k_a$ and $k_b$ the transverse 
momenta of the gluons at the ends of the propagator, and with kernel $K$,
\beq
K(k_a,k_b,k) = f_{\omega}(k_a+k,k_b) - {k_{a\perp}^2\over k_{\perp}^2 + 
(k_a+k)_{\perp}^2}\, f_{\omega}(k_a,k_b)\, ,\label{kern}
\eeq
where the first term accounts for the emission of a gluon of momentum
$k$ and the second for the virtual radiative corrections. It must be
noted, though, that eq.~(\ref{bfklb}) has been derived in the
{\sl multi-Regge kinematics}, which presumes that the produced gluons,
with momenta $p_a$, $p_b$ and $k$ are
strongly ordered in rapidity and have comparable transverse momenta 
\beq
\eta_a \gg \eta \gg \eta_b; \qquad |p_{a\perp}|\simeq|k_\perp|
\simeq|p_{b\perp}|\, ,\label{mrk}
\eeq
with $p_{a\perp} =- k_{a\perp}$, $p_{b\perp} = k_{b\perp}$. The
solution of eq.~(\ref{bfklb}), transformed from moment space to 
$\eta$ space, is
\bea
f(k_a,k_b,\eta)\, &=& \int {d\omega\over 2\pi i}\, e^{\omega\eta}\, 
f_{\omega}(k_a,k_b)\nn\\ &=& {1\over (2\pi)^2 k_{a\perp} k_{b\perp}} 
\sum_{n=-\infty}^{\infty} e^{in\tilde{\phi}}\, \int_{-\infty}^{\infty} d\nu\, 
e^{\omega(\nu,n)\eta}\, \left(k_{a\perp}^2\over k_{b\perp}^2
\right)^{i\nu}\, ,\label{solc}
\eea
with $\tilde\phi$ the azimuthal angle
between $k_a$ and $k_b$, $\eta\simeq\ln(\hat s/|\th|)\simeq
\ln(\hat s/\kt^2)$ the evolution parameter of the propagator,
with $\eta\gg 1$, and $\omega(\nu,n)$ the eigenvalue of
the BFKL equation whose maximum $\omega(0,0)=4\ln{2}N_c\a_s/\pi$
yields the known power-like growth of $f$ in energy~\cite{bal}.

In inclusive dijet production in hadron-hadron collisions
the resummed parton cross section for gluon-gluon scattering
is~\cite{mn},
\beq
{d\hat\sigma_{gg}\over dk_{a\perp}^2 dk_{b\perp}^2 d\phi} = 
{\pi N_c^2 \a_s^2 \over 2
k_{a\perp}^2 k_{b\perp}^2}\, f(k_a,k_b,\eta)\, ,\label{ppbar}
\eeq
with $\phi$ the azimuthal angle between the tagging jets, 
$\phi=\tilde{\phi}+\pi$. At the hadron level we convolute the parton 
cross section (\ref{ppbar}) with parton distribution functions ($pdf$),
$f(x,\mu_F^2)$. Thus, in the high-energy limit, and e.g. at fixed parton
momentum fractions, we have
\beq
{d\sigma\over dx_A dx_B dk_{a\perp}^2 dk_{b\perp}^2 d\phi}\, =\,
f_{eff}(x_A,\mu_F^2)\, f_{eff}(x_B,\mu_F^2)\,
{d\hat\sigma_{gg}\over dk_{a\perp}^2 dk_{b\perp}^2 d\phi}\, .\label{mnfac}
\eeq
with the effective $pdf$'s
\beq
f_{eff}(x,\mu_F^2) = G(x,\mu_F^2) + {4\over 9}\sum_f
\left[Q_f(x,\mu_F^2) + \bar Q_f(x,\mu_F^2)\right], \label{effec}
\end{equation}
with the sum over the quark flavors. In order to detect evidence
of a BFKL-type behavior, we'd like to see how $f(k_a^2,k_b^2,\eta)$
grows with $\eta$. In inclusive dijet production, $\eta$ is the rapidity 
difference between the tagging jets, $\Delta\eta=ln(\hat s/\kt^2)=\eta_1-
\eta_2$, and accordingly evidence of the BFKL dynamics is searched
in dijet events at large rapidity intervals~\cite{d0}. To obtain 
$\Delta\eta$ as large as possible, we minimize $k_{\perp}$, i.e.
the jet transverse energy, and maximize $\hat s$; since $\hat s = x_A
x_B s$ this may be achieved in two ways: 
\begin{itemize}
\item by fixing the parton momentum fractions $x$ and letting the hadron 
center-of-mass energy $s$ grow, and then measuring e.g. the dijet
production rate $d\sigma/ dx_A dx_B$ \cite{mn}. The theoretical advantage 
in this set-up is that $pdf$ variations
are minimised, while variations in the parton dynamics, and thus in
the eventual underlying BFKL behavior, are stressed. The experimental
drawback is of course that one needs different colliding-beam energies,
which have become just recently available with the colliding beams 
at Tevatron running at $\sqrt s = 630$~GeV, besides the usual data 
sample at $\sqrt s = 1800$~GeV. 
\item else, we may keep $s$ fixed and let the $x$'s grow. That is
experimentally much easier to realize (and before the 630 GeV run at the
Tevatron it was the only possibility), however, it is theoretically
unfavourable because as the $x$'s grow the $pdf$'s fall off, so
it is harder to disentangle the eventual BFKL-induced rise of the
parton cross section from the $pdf$'s fall off. We may resort then to
less inclusive observables: it was noted that the kinematic
correlation between the tagging jets, which to leading order 
are supposed to be back-to-back, is diluted as the rapidity distance 
$\Delta\eta$ grows. This is due to the more abundant gluon radiation 
between the tagging jets, which blurs the information on the mutual
position in transverse momentum space. Accordingly the transverse 
momentum inbalance \cite{DS}, and the azimuthal angle decorrelation 
\cite{DS}, \cite{stir} have been proposed as BFKL observables.
\end{itemize}

The BFKL theory, being a LL resummation and not an exact
calculation, makes a few approximations which, even
though formally subleading, may be important
for any phenomelogical purposes. I shall list them in
random order
\begin{itemize}
\item[i)] The BFKL resummation is performed at fixed coupling constant, 
thus any variation in its scale, $\a_s(\nu^2)=\a_s(\mu^2) - 
b_0\ln(\nu^2/\mu^2)\a_s^2(\mu^2) + \dots$, with $b_0= (11N_c-2n_f)/
12\pi$ and $n_f$ the number of quark flavors, would appear in the 
next-to-leading-logarithmic (NLL) terms, because it yields terms
of $O(\a_s^n\ln(\nu^2/\mu^2)\ln^{n-1}(\s/|\th|))$\footnote{Modulo 
pathological behaviors of the BFKL ladder as $\nu^2\ra s$, which
seems to generate LL terms which are not in the BFKL ladder \cite{bf}.}.
\item[ii)] From the kinematics of two-parton production at
$\s\gg|\th|$ we identify the rapidity interval between the tagging
jets as $\Delta\eta\simeq\ln(\hat s/|\th|)\simeq\ln(\hat s/\kt^2)$,
however, we know from the exact kinematics that $\Delta\eta =
\ln(\hat s/|\th|-1)$ and $|\th| = k_{\perp}^2 (1+\exp(-\Delta\eta))$,
therefore the identification of the rapidity interval $\Delta\eta$
with $\ln(\hat s/|\th|)$ is up to next-to-leading terms.
\item[iii)] Because of the strong
rapidity ordering any two-parton invariant mass is large. Thus there
are no collinear divergences in the LL resummation in the BFKL
ladder; jets are determined
only to leading order and accordingly have no non-trivial structure.
\item[iv)] Finally, energy-momentum is not conserved, and
since the momentum fraction $x$ of the incoming parton is
reconstructed from the kinematic variables of the outgoing partons,
the BFKL predictions may be affected by large numerical errors.
In particular, if $n+2$ partons are produced, we have
\beq
x_{A(B)} = {p_{a\perp}\over\sqrt{s}} e^{(-)\eta_a} +  
\sum_{i=1}^n {k_{i\perp}\over\sqrt{s}} e^{(-)\eta_i}
+ {p_{b\perp}\over\sqrt{s}} e^{(-)\eta_b}\, ,\label{nkin}
\eeq
where the minus sign in the exponentials of the right-hand side applies
to the subscript $B$ on the left-hand side. In the BFKL theory, the LL
approximation and the kinematics (\ref{mrk}) imply that
in the determination of $x_A$ ($x_B$) only the 
first (last) term in eq.~(\ref{nkin}) is kept, 
\bea
x_A^0 &=& {p_{a\perp}\over\sqrt{s}} e^{\eta_a}\, ,\nn\\
x_B^0 &=& {p_{b\perp}\over\sqrt{s}} e^{-\eta_b}\, .\label{nkin0}
\eea
The ensuing violation of
energy-momentum conservation and the neglected terms in 
eq.(\ref{nkin0}) are formally subleading. However, they may be important
for any phenomelogical purposes. Indeed, a comparison within dijet
production of the three-parton production to ${\o}(\a_s^3)$ with the
exact kinematics to the truncation of the BFKL ladder to
${\o}(\a_s^3)$ shows that the LL approximation may severely
underestimate the exact evaluation of the $x$'s (\ref{nkin}),
and therefore entail sizable violations of energy-momentum
conservation, even though the extent to which this is true
depends on the specific production rate considered \cite{DS2}.

The phenomenological improvement of the BFKL ladder through
feedback from the exact three-parton kinematics cures only
partially the pathological violation of energy-momentum 
conservation. A systematic approach 
is to require energy-momentum conservation at each
stage in the gluon emission in the BFKL ladder. This may
be achieved through a Monte Carlo implementation of the
BFKL equation (\ref{bfklb}) \cite{carl}, \cite{os}.
In particular, one can define a scale $\mu$, such that
$\mu \ll k_a, k_b$ and then split the real contribution to, 
i.e. the first term of, the BFKL kernel (\ref{kern}), into
unresolved (for $k < \mu$) and resolved (for $k > \mu$)
contributions; next, one combines the unresolved contribution,
for which $f_{\omega}(k_a+k,k_b) \simeq f_{\omega}(k_a,k_b)$,
with the virtual one, i.e. with the second term of eq.~(\ref{kern}),
in order to cancel the infrared singularities. One is then left over with
the integral over $k > \mu$ for the resolved contribution, which
is amenable to Monte Carlo resolution. Finally, one can check
that the precise choice of the scale $\mu$ is immaterial 
\cite{carl}, \cite{os}.
\end{itemize}

\subsection{Dijet production at fixed parton momentum fractions}
\label{sec:unouno}

Now we go back to the first scenario outlined above, i.e. increasing
the rapidity interval $\Delta\eta$ by letting $s$
grow at fixed $x$'s. We may compute the ensuing
cross section $d\sigma/ dx_A dx_B$
through the BFKL resummation from eq.~(\ref{mnfac}) 
by integrating over the jet transverse energies in various stages
of approximation, and consider the ratio $R$ of dijet production at
$\sqrt s_1 = 1800$~GeV and $\sqrt s_2 = 630$~GeV,
\beq
R(x_A,x_B;s_1,s_2) = {d\sigma(s_1)/ dx_A dx_B\over d\sigma(s_2)/ 
dx_A dx_B}\, ,\label{ratio}
\eeq
for values of $x_{A,B}$, and therefore of $\Delta\eta$, large enough
that BFKL is a sensible approximation to consider. Thus, we choose
$x_A=x_B=0.1$ for which $\Delta\eta(s_1)\simeq 4.5$ and 
$\Delta\eta(s_2)\simeq 2.3$, and $x_A=x_B=0.2$ for which 
$\Delta\eta(s_1)\simeq 5.9$ and $\Delta\eta(s_2)\simeq 3.7$.
\begin{itemize}
\item[a)] The simplest, and least accurate, approximation consists in 
fixing the renormalization scale of $\a_s$, the factorization scale of 
the $pdf$'s as well as the momentum transfer at the jet minimum 
transverse energy, $\mu_R^2=\mu_F^2=|\th|= k_{\perp min}^2$,
and in determining the parton momentum fractions through 
eq.~(\ref{nkin0}). Then the $pdf$'s factor
out of the integrals over transverse momentum which may be easily
performed analitically. Thus the cross section (\ref{mnfac}) becomes
\beq
{d\sigma\over dx_A^0 dx_B^0} = f_{eff}(x_A^0,k_{\perp min}^2)\, 
f_{eff}(x_B^0,k_{\perp min}^2)\, {\pi N_c^2 \a_s^2 \over 2
k_{\perp min}^2}\, f(\Delta\eta)\, ,\label{muel}
\eeq
with 
\bea
f(\Delta\eta) &=& \int_{-\infty}^{\infty} d\nu\, 
{e^{\omega(\nu,n=0) \Delta\eta}\over \nu^2+{1\over 4}} \nn\\ 
&\simeq& {e^{4\ln2 N_c \a_s \Delta\eta/\pi}\over 
\sqrt{7\zeta(3) N_c \a_s \Delta\eta/2}}\, ,\label{saddle}
\eea
with $\zeta(3) = 1.202...$ and where in the second line we have 
performed a saddle-point evaluation of the integral over $\nu$.
Accordingly, we get $R(x_A=x_B=0.1;s_1,s_2) = 1.66$ and 
$R(x_A=x_B=0.2;s_1,s_2) = 1.84$.
\item[b)] Else, we may fix the renormalization/factorization scales
and the momentum transfer in terms of the jet transverse energies,
e.g. we may choose $\mu_R^2=\mu_F^2=|\th|= k_{a\perp} 
k_{b\perp}$ \cite{DS}, and then integrate eq.~(\ref{mnfac})
numerically\footnote{Note then that $\a_s$ runs as a function
of $k_{a\perp}$ and $k_{b\perp}$ in eq.~(\ref{solc}), however, 
it is still fixed
within the BFKL ladder (\ref{bfklb}), i.e. it is not a function of
the transverse energies of the gluons emitted along the ladder.}.
For the sake of later comparison with the Monte Carlo results,
it is convenient to consider the averaged ratio $\bar R$, obtained
by integrating numerator and denominator of the right-hand side of 
eq.~(\ref{ratio}) over a range of $x_{A,B}$, e.g. over $0.1 \le
x_{A,B} \le 0.2$. One then finds $\bar R(0.1\le x_A,x_B\le 0.2;s_1,s_2)
\simeq 1.5$. As expected, the ratio is somewhat smaller than in
approximation $a)$ since in that instance the choice
$\mu_R^2=\mu_F^2=|\th|= k_{\perp min}^2$ overestimates the $pdf$'s, 
the coupling constant and the size of the rapidity interval.
\item[c)] Finally, the averaged ratio $\bar R$ may be computed
using a Monte Carlo evaluation of eq.~(\ref{bfklb}), 
with energy-momentum conservation
in the gluon emission along the BFKL ladder \cite{carl}, \cite{os}
\footnote{The Monte Carlo event generator of ref.~\cite{os} includes
also the option of running $\a_s$ within the BFKL ladder.}.
It is possible to achieve that through a two-stage process, namely
requiring energy-momentum conservation only on the kinematic part
of the cross section (\ref{mnfac}), while still using the approximate
parton momentum fractions (\ref{nkin0}) in determining $\s$ in the
squared amplitudes, or else using the exact parton momentum fractions 
(\ref{nkin}) everywhere in the cross section (\ref{mnfac}) \cite{DS2}.
The results of both these methods, though, agree with one another
and with the one of approximation $b)$, within the statistical error, 
i.e. $\bar R(0.1\le x_A,x_B\le 0.2;s_1,s_2) \simeq 1.5$ \cite{priv}. That
entails apparently that energy-momentum conservation is not so relevant 
for the averaged ratio $\bar R$, however this conclusion is misleading
since the absolute cross section $d\sigma/dx_A dx_B$ evaluated through the BFKL
ladder increases with respect to the ${\o}(\a_s^2)$ i.e. the lowest
order calculation within approximation $b)$, while it decreases with 
respect to the ${\o}(\a_s^2)$ calculation in
the Monte Carlo evaluation \cite{priv}. In addition, the averaged 
ratio is different in the approximation $b)$ and in the Monte Carlo 
evaluation if other values of $x_A,x_B$ and/or of $s_1,s_2$ (like
e.g. $\sqrt s_1 = 14$~TeV and $\sqrt s_2 = 1800$~GeV) are used
\cite{priv}. Thus the insensitivity of $\bar R(0.1\le x_A,x_B\le 
0.2;s_1,s_2)$ to the implementation of energy-momentum conservation
in the BFKL ladder seems to be an accident of the values
of $x_A,x_B;s_1,s_2$ chosen above.
\end{itemize} 
It remains to be seen what the result is for the ratio (\ref{ratio})
if a standard Monte Carlo event generator, like HERWIG \cite{her}, 
that includes 
Altarelli-Parisi-evolved parton showers and coherence, but no BFKL
evolution, is used\footnote{Before any comparison is made, it should 
be stressed that none of the BFKL Monte Carlo event generators and 
calculations described above includes hadronization effects.}.

\subsection{Dijet production at fixed hadron energy}
\label{sec:unodue}

Let us consider now the second scenario outlined above, namely 
increasing the rapidity interval $\Delta\eta$ by letting $\hat s$
grow at fixed $s$. That entails an increase in the $x$'s, thus
it is more convenient to measure the inclusive dijet rate at
fixed rapidities $d\sigma/d\eta_1 d\eta_2$ (or equivalently,
at fixed $\Delta\eta$ and rapidity boost $\bar\eta = (\eta_1 +\eta_2)
/2$). The ensuing factorization formula in the high-energy limit is
\beq
{d\sigma\over d\Delta\eta d{\bar\eta} dk_{a\perp}^2 
dk_{b\perp}^2 d\phi}\, =\, x_A f_{eff}(x_A,\mu_F^2)\, x_B 
f_{eff}(x_B,\mu_F^2)\, {d\hat\sigma_{gg}\over dk_{a\perp}^2 
dk_{b\perp}^2 d\phi}\, .\label{dsfac}
\eeq
Since the $x$'s grow linearly with the transverse momenta 
(cf. eq.~(\ref{nkin}-\ref{nkin0})), when integrating over the 
latter, the BFKL-induced growth of the dijet 
rate is upset by the falling parton luminosities. This is even
more noticeable for the dijet rate calculated with the BFKL ladder 
than for the rate computed to lowest order, i.e. ${\o}(\a_s^2)$,
because the additional gluon radiation generated by the BFKL ladder
makes the dijet rate to run out
of phase space more rapidly than the one computed to lowest 
order \cite{DS}, \cite{stir}, \cite{vitt}. Accordingly, it was
proposed to measure the transverse momentum inbalance \cite{DS}, 
and the decorrelation in the azimuthal angle $\phi$ between the
tagging jets \cite{DS}, \cite{stir},
as signatures of an eventual BFKL evolution. The azimuthal angle 
decorrelation turns out to be easier to measure \cite{d0}, so we 
shall concentrate on that. The inclusive distribution
$d\sigma/ d\phi$ is centered around the peak $\phi = \pi$ since
to lowest order kinematics require the jets to be back-to-back,
$d\sigma/ d\phi \sim \delta(\phi-\pi)$; the additional gluon
radiation, induced by parton showers and hadronization,
smears the $\delta$ function into a bell curve peaked at $\phi = \pi$.
However, if we look at the distribution also as a function of the
rapidity distance between the jets, $d\sigma/d\Delta\eta d\phi$,
we expect that the larger $\Delta\eta$ the larger the smearing
of the distribution \cite{DS}, \cite{stir}. The reason is that the
BFKL-induced gluon radiation, which is roughly constant
per unit of rapidity, is so more abundant for a larger
$\Delta\eta$, and accordingly the information on the mutual
position of the jets in transverse momentum space is more diluted as
$\Delta\eta$ grows. This has been experimentally confirmed \cite{d0}.
From the calculational point of view,
it is easier to evaluate the moments of $\phi$ \cite{stir},
\beq
<\cos n(\phi\!-\!\pi)>\, = 
{\int_0^{2\pi} d\phi \cos n(\phi\!-\!\pi)\, 
\big(d\sigma/d\Delta\eta d\phi\big)
\over \int_0^{2\pi} d\phi \big(d\sigma/d\Delta\eta d\phi\big)}. 
\eeq
For a $\delta$-function distribution at $\phi=\pi$, as occurs at the lowest
order, all of the moments will equal one, while for a flat distribution all
of the moments will equal zero for $n\ge1$.  Thus, the decay of the moments 
from unity is a good measure of the decorrelation in $\phi$.

The decorrelation of the first moment has been measured for values of
the rapidity interval up to $\Delta\eta = 5$ and for transverse
momenta of the jets $k_{a\perp} = 50$~GeV and $k_{b\perp} = 20$~GeV
\cite{d0}, and preliminary results are available from a larger data
sample for rapidity intervals up to $\Delta\eta = 6$ and for
a symmetric configuration $k_{a\perp} = k_{b\perp} = 20$~GeV \cite{Jun}. 

The evaluation of $<\cos (\phi\!-\!\pi)>$ using the BFKL ladder
without energy-momentum conservation (approximation $b)$ of 
sect.~\ref{sec:unouno}) yields too much decorrelation between
the jets \cite{stir}, \cite{DS3}, as compared to the data \cite{d0}.
This is mainly due to the parton momentum fractions (\ref{nkin0})
sizeably underestimating the exact kinematics (\ref{nkin}),
and thus violating substantially energy-momentum conservation.
As hinted in item $iv)$ of sect.~\ref{sec:unozero}, a phenomenological
improvement may be achieved by noting that since the identification 
$\Delta\eta\simeq\ln(\hat s/|\th|)$ holds up to next-to-leading terms,
and the difference between the exact (\ref{nkin}) and the approximate
(\ref{nkin0}) kinematics resides also in the next-to-leading terms,
we may use the three-parton production with the exact kinematics, 
i.e. eq.~(\ref{nkin}) 
with $n=1$, to define an effective rapidity interval $\Delta\hat\eta$,
such that if we replace $\Delta\eta\ra\Delta\hat\eta$ in the BFKL 
ladder and truncate it to ${\o}(\a_s^3)$ we reproduce the exact 
three-parton contribution to dijet production \cite{DS2}.
Using the effective rapidity interval $\Delta\hat\eta$ in the BFKL 
ladder \cite{DS3} improves sizeably the agreement with the data 
\cite{d0}, even though the BFKL ladder shows still too much
decorrelation. Eventually the best solution is to require 
energy-momentum conservation for the emission of each gluon along the 
BFKL ladder, through a Monte Carlo evaluation of eq.~(\ref{bfklb}) 
\cite{carl}, \cite{os}. Accordingly the first
moment $<\cos (\phi\!-\!\pi)>$ has been evaluated \cite{os}
using the kinematic cuts of the more recent D0 analysis
\cite{Jun}. The BFKL Monte Carlo generator still seems to yield too much
decorrelation as compared to the data \cite{Jun}, even though
it is possible to notice in the Monte-Carlo-generated curve the
workings of energy-momentum conservation as the boundary of phase
space is approached, i.e. for the largest values of $\Delta\eta$
kinematically attainable. However, it is
premature to draw conclusions since the D0 data are preliminary
and, as noted at the end of sect.~\ref{sec:unouno}, the BFKL
Monte Carlo generators do not include hadronization effects.

It must also be stressed that the HERWIG Monte Carlo generator 
\cite{her} is in perfect agreement with the data for the first
moment $<\cos (\phi\!-\!\pi)>$, which entails that the azimuthal
angle decorrelation 
is not exclusive to the gluon radiation induced by the BFKL
ladder, but can be found also, and with better agreement, in 
standard patterns of gluon radiation.

\subsection{Conclusions for dijet production}
\label{sec:unotre}

The interplay between data and theory has allowed us to improve
considerably the phenomenological predictions based on the BFKL
ladder, however, we do not see yet, in the context of dijet production
at the Tevatron, a clear signal of BFKL behavior. On one hand,
the several approximations listed in sect.~\ref{sec:unozero} and
embodied in the LL resummation, on which the BFKL ladder is based, 
seem somewhat unreliable for the kinematic range available in
dijet production at the Tevatron; to
this effect a full NLL calculation, which is not available yet 
\cite{DD}, should achieve a considerable improvement. On the other
hand, the comparison with experiment has taught us that it may be
not so easy to disentangle the effects of BFKL-induced gluon
radiation from the ones of the more standard and well understood
Altarelli-Parisi gluon radiation, based on collinear emission.
In this search other eventual BFKL observables, like the transverse 
energy in the central rapidity region between the tagging jets 
\cite{carl}, or the single particle $k_{\perp}$ spectra \cite{kuhl},
as well as a deeper analysis of the existing ones,
e.g. exploring moments higher than the first in the azimuthal angle
decorrelation, should be sought after.

\section{Forward jet production in DIS}
\label{sec:due}
\subsection{Fixed-order calculations}
\label{sec:dueuno}

A variant of dijet production at hadron colliders is the production
in DIS of a jet close in rapidity to the proton remnants, and with
transverse momentum comparable to the virtuality of the photon
$k_{\perp}^2 \simeq Q^2$, in order to fulfil as closely as possible 
the constraints of the multi-Regge kinematics (\ref{mrk}) 
\cite{aldis}, \cite{foll}. In this case the resummation
parameter $\eta$ of the BFKL ladder (\ref{solc}) is $\eta =ln(\hat
s_{\gamma p}/k_{\perp}^2) \simeq ln(\hat s_{\gamma p}/Q^2) \simeq
ln(x/x_{bj})$, with $\s_{\gamma p}$ the center-of-mass energy between the
photon and the struck parton, $x_{bj}=Q^2/s$ the Bjorken variable
($s$ is the lepton-proton energy) and $x$ the momentum fraction of 
the struck parton. In order to have $\eta\gg 1$, we require that 
$x\gg x_{bj}$.

To clarify the picture of forward jet production in DIS,
let us start from the parton model, where a quark is struck by the
virtual photon. In this case the kinematics costrain $x$ to be 
$x = x_{bj}$, and a jet may form in the photon direction, usually termed the
current fragmentation region. However, since we have required 
$x\gg x_{bj}$, there is no contribution to forward jet production in DIS
from the parton model (just like for the structure function $F_L$). 

The leading-order (LO) contribution to forward jet production comes
to ${\o}(\a_s)$ from tree-level production of two partons.
Then it is possible to compute either LO two-jet 
production, one of which required to be forward, or NLO forward-jet 
production; however requiring the jet to be forward makes the latter
effectively LO since the one-loop one-parton production, which has
$x = x_{bj}$, is kinematically forbidden. Therefore we expect
that to ${\o}(\a_s)$ the two-jet (one forward) rate and the
forward-jet rate are of the same size, since in the
latter we are simply integrating over the kinematic variables 
of the extra jet, produced in the current fragmentation region.

To ${\o}(\a_s^2)$ we may produce three final-state partons at tree
level,with the novel feature of
gluon exchange in the cross channel, which in the high-energy limit
$\s\gg\th$ dominates the production rate; thus the LO three-jet (one
forward) rate turns out to be of the same size as the LO two-jet 
rate, even though it is one order higher in $\a_s$,
and therefore in general expected to be an order of
magnitude smaller \cite{mz}. Accordingly, the ${\o}(\a_s^2)$ NLO two-jet 
(one forward) rate, which is computed from 
tree-level three-parton and one-loop two-parton final states,
turns out to be sizeably bigger than the ${\o}(\a_s)$ LO two-jet rate
\cite{mz}, because the latter does not have gluon exchange in the cross 
channel. Finally, since two-loop one-parton production, which has
$x = x_{bj}$, is kinematically forbidden, it is also feasible
to compute the NNLO forward-jet rate, because requiring the jet 
to be forward makes the rate effectively NLO; analogously to the
${\o}(\a_s)$ case we expect it to be of the same size as NLO 
two-jet production. In addition, since the forward-jet rates
above are dominated by gluon exchange in the cross channel,
which to ${\o}(\a_s^2)$ occurs in tree-level diagrams, the 
dependence on the factorization/renormalization scales 
is sizeable \cite{mz}, even though the rates have been computed 
at NLO.

Since no new kinematic features appear to ${\o}(\a_s^3)$, we
expect that the LO four-jet (one forward) rate
should be markedly smaller than the LO three-jet one,
while NLO and LO three-jet rates should be of the same order 
of magnitude, however, in the ${\o}(\a_s^3)$ calculation the 
dependence on the factorization/renormalization scales should
be sensibly reduced. 

\subsection{The BFKL ladder}
\label{sec:duedue}

The BFKL ladder builds up the radiative corrections to
gluon exchange in the cross channel, thus it appears to
${\o}(\a_s^2)$ in forward-jet production. The corresponding
lepton-parton cross section is obtained by convoluting in $k_{\perp}$
space the BFKL ladder (\ref{solc}) with the coefficient function
$\gamma^* g^* \ra q\bar q$, with off-shell gluon and photon,
\beq
{d\hat \sigma\over dx_{bj} dQ^2 dk_{\perp}^2 d\phi} = \sum_q e_q^2 {N_c\a^2 
\a_s^2 \over \pi^2 (Q^2)^2 k_{\perp}^2 x_{bj}}
\int {dv_{\perp}^2\over v_{\perp}^2} f(v_{\perp},k_{\perp},\eta) 
{\cal F}(v_{\perp},Q^2,y)\, ,\label{hot}
\eeq
with $y = Q^2/x_{bj} s$ the electron energy loss, ${\cal F}$ the off-shell
coefficient function, $k_{\perp}$ and $v_{\perp}$ respectively 
the transverse momenta of the 
forward jet and of the gluon attaching to the $q\,\bar q$ pair,
and with the sum over the quark flavors in the $q\,\bar q$ pair.
After substituting the off-shell coefficient function ${\cal F}$,
the integral over $v_{\perp}$ on the right-hand side of 
eq.~(\ref{hot}) may be performed, yielding \cite{us},
\bea
& & \int {dv_{\perp}^2\over v_{\perp}^2} f(v_{\perp},k_{\perp},\eta) 
{\cal F}(v_{\perp},Q^2,y) \nn\\ & & = {\pi\over 8} 
\left({Q^2\over k_{\perp}^2}\right)^{1/2}
\int_0^{\infty} d\nu\, \cos\left(\nu\ln{Q^2\over k_{\perp}^2}\right)\, 
{\sinh(\pi\nu)\over\cosh^2(\pi\nu)}\, {1\over\nu(1+\nu^2)} \nn\\ 
& & \times \left(e^{\omega(\nu,0)\eta}\,\left[{1\over2}
\left(\nu^2+{9\over 4}\right)\left[1+(1-y)^2\right] +
2 \left(\nu^2+{1\over 4}\right)(1-y) \right] \right.\nn\\
& & - \left. e^{\omega(\nu,2)\eta}\, \cos(2\phi)\, \left(\nu^2+{1\over 4}\right)
(1-y)\right)\, .\label{jet}
\eea
In the high-energy limit, and at fixed parton momentum fraction, the
forward-jet rate is,
\beq
{d\sigma\over dx_{bj} dQ^2 dx dk_{\perp}^2 d\phi} = f_{eff}(x,\mu_F^2)
{d\hat \sigma\over dx_{bj} dQ^2 dk_{\perp}^2 d\phi}\, ,\label{xfix}
\eeq
with the parton cross section (\ref{hot}). By
producing the jet forward, we make $x$ large;
$\eta$ is then made larger and larger by making 
$x_{bj}$ smaller and smaller. Thus the advantage of forward-jet 
production in DIS is that a fixed-energy $ep$ collider is nonetheless 
a variable-energy collider in the photon-proton frame \cite{aldis},
and we may realize the first scenario of sect.~\ref{sec:unozero}
(fixed $x$, variable $\s$) with a large range, in principle a
continuum, of energies, instead of just two.

The H1 Collaboration at HERA \cite{H1} has measured the forward-jet 
rate with $x > 0.025$, $2\cdot 10^{-4} <x_{bj}< 2\cdot 10^{-3}$
(thus with typical values of $2.5 < \eta <4.8$), $5\,{\rm GeV}^2 <
Q^2 < 100\,{\rm GeV}^2$, $k_{\perp} >5$~GeV and $0.5 < k_{\perp}^2/Q^2 
< 4$. The ${\o}(\a_s^2)$ forward-jet rate
mentioned in sect.~\ref{sec:dueuno}, based on the NLO Monte Carlo
program MEPJET \cite{mep}, is substantially larger than 
the ${\o}(\a_s)$ one, as expected, however, it falls below 
the data by a factor 4$\div$5 \cite{mz}, which hints that the radiative 
corrections beyond ${\o}(\a_s^2)$ are important. 
A calculation of the forward-jet rate through the BFKL ladder,
performed like in approximation $b)$ of sect.~\ref{sec:unouno}, 
i.e. by fixing $\mu_R^2=\mu_F^2=k_{\perp}^2$ and integrating 
eq.~(\ref{xfix}) numerically, shows a good agreement with the data
\cite{us}. On the other hand, the forward-jet rate computed by
truncating the BFKL ladder to its lowest order, i.e. to
${\o}(\a_s^2)$ \cite{us}, is in good agreement with the exact 
${\o}(\a_s^2)$ forward-jet rate \cite{mz}, confirming that
in the high-energy limit the exact ${\o}(\a_s^2)$ calculation is 
dominated by gluon exchange in the cross channel, which forms the 
lowest-order contribution to the BFKL ladder.

A more recent and larger data sample from the H1 Collaboration 
\cite{prelh}, with $x > 0.035$, $5\cdot 10^{-4} <x_{bj}< 3.5\cdot 10^{-3}$
(thus with typical values of $2.3 < \eta <4.2$), $k_{\perp} >3.5$~GeV 
and $0.5 < k_{\perp}^2/Q^2 < 2$, basically confirms the earlier findings
\cite{H1}, namely the forward-jet rate computed through the NLO 
Monte Carlo program DISENT \cite{cat} falls well below the data;
while the rate based on the BFKL calculation \cite{us} fares better
but tends to overshoot the data, at the lower values of $x_{bj}$. 
However, a caveat is in order: DISENT and the BFKL calculation do not include
hadronization, while a simulation through the LEPTO Monte Carlo 
event generator \cite{ing}, based on Altarelli-Parisi-evolved 
parton showers, shows that hadronization effects are important.
It is also encouraging that the ARIADNE Monte Carlo generator 
\cite{leif}, which does not have the typical Altarelli-Parisi-induced
$k_{\perp}$ ordering but rather privileges comparable transverse
momenta in accordance with the multi-Regge kinematics (\ref{mrk}),
is very close to the data.

Analogously, the ZEUS Collaboration has measured the forward-jet rate
\cite{prelz}, with $x > 0.035$, $5\cdot 10^{-4} <x_{bj}< 5\cdot 10^{-3}$
(thus with typical values of $1.9 < \eta <4.2$), $k_{\perp} > 5$~GeV 
and $0.5 < k_{\perp}^2/Q^2 < 4$, and the same considerations done
for ref.~\cite{prelh} about the comparison with the different 
theoretical or phenomenological models apply here.

\subsection{Conclusions for forward-jet production}
\label{sec:duetre}

The analyses of the H1 and ZEUS Collaborations at HERA seem to 
indicate that forward-jet production is a good candidate as a
BFKL observable. In order to improve the agreement with the
data, and enforce energy-momentum conservation, a BFKL Monte Carlo 
generator \cite{carl}, \cite{os} should be used. However, some 
experimental issues like
the hadronization effects mentioned above or the excess in the
dijet rate as a function of $x_{bj}$ and $Q^2$ \cite{prelh}, 
\cite{prelz}, \cite{mart}, which might imply a contamination
from resolved-photon production, particularly at low $Q^2$,
should be understood before firm conclusions on forward-jet 
production may be reached.

The study of other BFKL observables in DIS, like the single
particle $k_{\perp}$ spectra \cite{kuhl} mentioned above, or,
in analogy to the analysis of sect.~\ref{sec:unodue},
the decorrelation of the azimuthal angle between the lepton
and the jet, which shows a peculiar shift of the distribution 
maximum from $\phi=\pi$ at large $x_{bj}$ to $\phi=\pi/2$ at
small $x_{bj}$ \cite{us}, \cite{vdd}, \cite{bdw}, should also 
be pursued.

For the future, the exchange of a BFKL ladder might also be studied
in the same theoretical framework as originally suggested in
ref.~\cite{bal}, namely in virtual photon-photon scattering 
in $e^+e^-$ collisions \cite{haut}. This would provide a particularly clean
environment, benefitting from the absence of QCD initial-state 
radiation.

{\un {\sl Acknowledgements}} I wish to thank Lynne Orr, Carl Schmidt
and James Stirling for useful discussions about the Monte Carlo event
generators, the organizers of the 
XIIth Hadron Collider Physics symposium for the support, and the ITP
of SUNY at Stony Brook for the hospitality.

\end{document}